\documentclass[aps, twocolumn, showpacs, prl, letterpaper]{revtex4}

\usepackage{amsmath}
\usepackage{amssymb}
\usepackage{graphicx}
\usepackage{xspace}
\usepackage{upgreek}
\usepackage{accents}

\newcommand{\eg}{{e.g.,\/}\xspace}
\newcommand{\ie}{{i.e.,\/}\xspace}
\newcommand{\etal}{{\it et~al\/}\xspace}

\newcommand{\eq}[1]{(\ref{#1})}
\newcommand{\Eq}[1]{Eq.~(\ref{#1})}
\newcommand{\Eqs}[1]{Eqs.~(\ref{#1})} 
\newcommand{\Fig}[1]{Fig.~\ref{#1}} 
\newcommand{\Ref}[1]{Ref.~\cite{#1}}
\newcommand{\Refs}[1]{Refs.~\cite{#1}}

\newcommand{\mc}[1]{\mathcal{#1}}
\newcommand{\mcc}[1]{\mathfrak{#1}}
\newcommand{\msf}[1]{\mathsf{#1}}

\newcommand{\favr}[1]{\langle #1 \rangle}
\renewcommand{\vec}[1]{{\boldsymbol{\rm #1}}}
\newcommand{\oper}[1]{\hat{\vec{#1}}}
\newcommand{\pd}{\partial}

\newcommand{\kpt}[1]{{\kern #1 pt}}

\newcommand{\ds}{\displaystyle}
\newcommand{\msection}[1]{\textit{#1.}\ ---\ } 

\begin{document}

\title{Nonlinear dispersion of stationary waves in collisionless plasmas}
\author{I.~Y. Dodin and N.~J. Fisch}
\affiliation{Department of Astrophysical Sciences, Princeton University, Princeton, New Jersey 08544, USA}

\begin{abstract}
A nonlinear dispersion of a general stationary wave in collisionless plasma is obtained in a non-differential form from a single-particle oscillation-center Hamiltonian. For electrostatic oscillations in nonmagnetized plasma, considered as a paradigmatic example, the linear dielectric function is generalized, and the trapped particle contribution to the wave frequency shift $\Delta \omega$ is found analytically as a function of the wave amplitude $a$. Smooth distributions yield ${\Delta \omega \sim a^{1/2}}$, as usual. However, beam-like distributions of trapped electrons result in different power laws, or even a logarithmic nonlinearity, which are derived as asymptotic limits of the same dispersion relation.
\end{abstract}

\pacs{52.35.-g, 52.35.Mw, 52.25.-b, 45.20.Jj}


\maketitle

\msection{Introduction} Nonlinear stationary waves, such as Bernstein-Green-Kruskal (BGK) modes, remain of continuing interest \cite{refm:bgk, ref:schamel00}, including recently in connection with Raman backscattering \cite{ref:yampolsky09a} and new methods of phase space manipulation \cite{refm:appl}. However, essential properties of these waves are not apparent, because the waves are derived directly from the Vlasov-Maxwell system. The nonlinear dispersion relations (NDR) are obtained then in a differential form \cite{ref:schamel00, refm:ndr}, which is both specific to particular settings and may be analytically intractable, thus obscuring the underlying physical picture.

Here, we offer a universal non-differential NDR [\Eq{eq:ndr0}] with a transparent physical meaning. The new NDR reveals that the nonlinear properties of a wave in collisionless plasma are entirely determined by one function, namely, the single-particle oscillation-center (OC) Hamiltonian $\mc{H}$ \cite{foot:disp}. Once $\mc{H}$ is found, one can study the nonlinear effects systematically and hence keep track of effects that are easy to miss in \textit{ad hoc} calculations. Electrostatic waves in nonmagnetized plasma are considered as a paradigmatic example. For those, we show how various types of kinetic nonlinearities, previously known from different contexts, and also a new logarithmic nonlinearity are derived as asymptotic limits of the same dispersion relation [\Eq{eq:drpw}]. Besides that, the fundamental linear dielectric function is generalized [\Eqs{eq:epsilonF} and \eq{eq:eplin}], and the friction drag on trapped particles is predicted to affect the wave frequency sweeping in collisional plasmas.

\msection{Basic equations} To start, consider the plasma Lagrangian $L_\Sigma = L_{\rm em} + \sum_i L_i$, where $L_{\rm em}$ is the electromagnetic field Lagrangian, and $L_i$ are the Lagrangians of individual particles, also accounting for the interaction with the field. The plasma adiabatic dynamics on time scales large compared to the period of any oscillations in the system is then governed by the time-averaged Lagrangian, $\msf{L}_\Sigma = \favr{L_\Sigma}_t$ \cite{ref:whitham65}. Notice further that, in a stationary wave, particles can be described by some generalized canonical coordinates $\vec{\mc{Q}}_i$ and momenta $\vec{\mc{P}}_i$, referred to as OC variables \cite{foot:oc}, such that $\dot{\vec{\mc{Q}}}_i$ and $\vec{\mc{P}}_i$ remain constant. Since the particle dynamics is trivial in these variables, let us exclude them as separate degrees of freedom. This is done using Routh reduction \cite{my:mneg, book:arnold06}, which yields the Lagrangian $\msf{L} = \msf{L}_\Sigma - \sum_i \vec{\mc{P}}_i \cdot \dot{\vec{\mc{Q}}}_i$ that describes the wave only. Further, since
\begin{gather}\label{eq:ham}
\vec{\mc{P}}_i\cdot \dot{\vec{\mc{Q}}}_i - \favr{L_i}_t = \mc{H}_i
\end{gather}
is $i$th particle OC Hamiltonian \cite{my:mneg}, one obtains $\msf{L} = \favr{L_{\rm em}}_t - \sum_i \mc{H}_i$ \cite{foot:brizard}. Hence, the wave Lagrangian $\msf{L}$ per unit spatial volume is given by $\mcc{L} = \mcc{L}_{\rm em} - \sum_s n_s \favr{\mc{H}_s}$, where $\mcc{L}_{\rm em} = \favr{E^2-B^2}_{x,t}/(8\pi)$ (with averaging performed over both time and space), $\vec{E}$ and $\vec{B}$ are the electric and magnetic fields, summation is taken over different species $s$, $n_s$ are the corresponding space-average densities, and $\favr{\mc{H}_s}$ is the OC energy averaged over $\vec{\mc{P}}_s$. 

Assuming the wave spatial profile is prescribed, the dynamics of the wave is fully characterized by its amplitude $a$ (arbitrarily normalized) and canonical phase $\xi$; by definition, the latter increases at some \textit{constant} rate $\dot{\xi} \equiv \omega$, by $2\pi$ per the oscillation period $T = 2\pi/\omega$. Hence, $\mcc{L} = \mcc{L}(a, \dot{\xi})$, where we used that $\mcc{L}$ cannot depend on $\xi$ explicitly for it describes the dynamics on time scales $t \gg T$; cf. \Ref{ref:whitham65}. In particular, varying $\msf{L}$ with respect to $a$ at fixed $\omega$ yields $\pd_a \mcc{L} = 0$ \cite{foot:action}, or
\begin{gather}\label{eq:ndr0}
 \frac{1}{8\pi}\,\frac{\pd}{\pd a}\favr{E^2-B^2}_{x,t} - \sum_s n_s \frac{\pd\favr{\mc{H}_s}}{\pd a} = 0.
\end{gather}
Complemented by \Eq{eq:ham} for $\mc{H}_i$ \cite{foot:hl}, \Eq{eq:ndr0} is the sought NDR, with advantages that it (i) applies to any stationary wave in collisionless plasma, (ii) has a non-differential form, (iii) is nonperturbative in the field amplitude, (iv) allows understanding the wave properties by studying just $\mc{H}_s$, (v) is comprised of terms with transparent physical meaning. Below, examples are given that illustrate the power of this main result.

First, revisit linear waves, in which case there clearly \textit{must} exist modes of the form $\vec{E}, \vec{B} \propto e^{i\vec{k}\cdot \vec{x}}$; then the commonly known dispersion relation for linear waves without trapped particles \cite{book:stix} should follow. To confirm this, substitute $\mc{H}_s$ in the dipole approximation \cite{my:kchi}, namely, 
\begin{gather}\label{eq:ponder}
\mc{H}_s = \mc{H}_s^{(0)} + \Phi_s, \quad \Phi_s = -\vec{E}^*\cdot\oper{\alpha}_s\cdot\vec{E}/4,
\end{gather}
where $\mc{H}_s^{(0)}$ is some function of $\vec{\mc{P}}$ (and static fields, if any), $\Phi_s$ is the ponderomotive potential, and $\oper{\alpha}_s$ is the particle linear polarizability. Take $\vec{E} = a\vec{e}$, where $\vec{e}$ determines polarization; then $\pd_a \mc{H}_s = -\frac{1}{2}\, (\vec{e}^*\cdot\oper{\alpha}_s\cdot\vec{e})a$ and also $B = |\vec{n} \times \vec{e}| a$, where $\vec{n} \equiv c\vec{k}/\omega$, and $c$ is the speed of light. Hence, \Eq{eq:ndr0} gives $(\vec{e}^* \cdot \oper{\epsilon} \cdot \vec{e}) - |\vec{n} \times \vec{e}|^2 = 0$, where $\oper{\epsilon} \equiv 1 + \sum_s 4\pi n_s \favr{\oper{\alpha}_s}$ equals the linear dielectric tensor. Thus, our result coincides with the known dispersion relation \cite{book:stix} at prescribed $\vec{e}$, whereas the equation for $\vec{e}$ also can be recovered, by varying $\msf{L}$ with respect to $\vec{e}^*$.

\msection{Electrostatic wave} Now let us apply \Eq{eq:ndr0} to derive the dispersion of a nonlinear electrostatic wave in nonmagnetized plasma. Assume that ions are fixed; hence, only electron motion will be addressed, and the species index $s$ is dropped. Also, neglect fluid nonlinearities, which are of higher order in $a$ than the kinetic nonlinearities discussed below. Then, treating the wave as monochromatic is anticipated to yield asymptotically precise description at small amplitudes \cite{ref:rose01, ref:winjum07}. We hence introduce the wavenumber $k$ and the phase velocity $u = \omega/k$. (Both $u$ and electron velocities will be assumed nonrelativistic.) From \Eq{eq:ham}, it is seen then that $\mc{H}$ is conserved to a $u$-dependent term $\Delta \mc{H}$ when transferring from the laboratory frame $K$ to the reference frame $K'$ where the wave field is static. Since $\Delta \mc{H}$ is independent of $a$, for the purpose of using \Eq{eq:ndr0} it only remains to find $\mc{H}$ in $K'$, which is done as follows.

\begin{figure}[t]
\centering
\includegraphics[width=.48\textwidth]{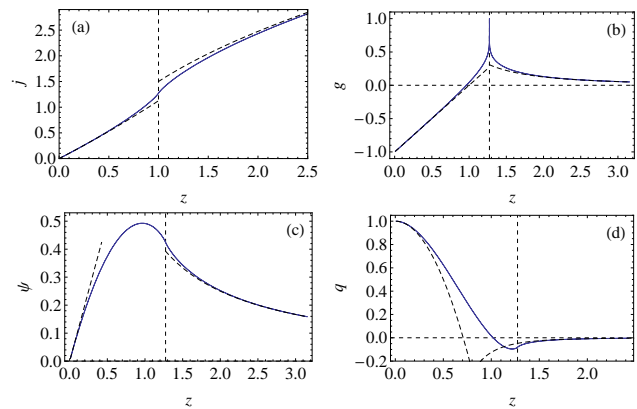}
\caption{Auxiliary dimensionless functions (solid): (a) $j(z)$, (b) $g(z)$, (c) $\psi(z)$, (d) $q(z)$. The vertical dashed lines show where the functions are nonanalytic. Also shown are asymptotes and asymptotic approximations (dashed) flowing from \Eq{eq:gasym}, except in (a), where the approximations used are $j(z \ll 1) \approx z + z^2/8$ and $j(z \gg 1) \approx 2z^{1/2} - (2z^{1/2})^{-1}$.}
\label{fig:fig}
\end{figure}

First consider the electron true Hamiltonian in $K'$,
\begin{gather}\label{eq:hamr}
H(x,p) = p^2/(2m) + e\varphi_0 \cos (kx),
\end{gather}
where $m$ and $e$ are the particle mass and charge, $p = m(v - u)$ is the corresponding momentum ($v$ being the velocity in $K$), and, for clarity, the amplitude of the potential energy is defined such that $a \equiv k^2e\varphi_0/(m\omega^2) > 0$. Governed by \Eq{eq:hamr}, both passing and trapped particles will undergo oscillations which are convenient to describe in terms of the action $J \propto \oint p\,dx $ and the conjugate canonical phase $\theta$, which will serve as $\vec{\mc{P}}$ and $\vec{\mc{Q}}$ in this case. Specifically, choose the coefficient in the expression for the passing-particle action such that $J = |p|/k$ for large $p$ (we assume $k>0$), and, for trapped particles, such that $J$ is continuous across the separatrix. Then, $J = \hat{J}a^{1/2}j(r)$, where $\hat{J} = m\omega/k^2$, and [\Fig{fig:fig}(a)]
\begin{gather}\label{eq:j}
j(r) = \frac{4}{\pi}\times
\left\{ 
\begin{array}{ll}
\ds \msf{E}(r) + (r-1) \msf{K}(r), & \quad r < 1,\\[5pt]
\ds r^{1/2}\msf{E}(r^{-1}), & \quad r > 1
\end{array} 
\right.
\end{gather}
is a continuous function of the normalized energy $r \equiv (H + e\varphi_0)/(2e\varphi_0)$, such that $j = 0$ for a particle resting at the bottom of the potential trough ($r = 0$), with the corresponding value at the separatrix ($r = 1$) being $j_* = 4/\pi$. (Here $\msf{K}$ and $\msf{E}$ are the complete elliptic integrals of the first and second kind, respectively; cf., \eg \Ref{ref:best68}.)

Since the generating function of the canonical transformation $(x,p) \to (\theta, J)$ clearly does not depend on time explicitly, one gets $\mc{H}(J) = H (x,p)$, or $\mc{H}(J, a) = (2ar - a)\hat{J}\omega$, where the dependence on $a$ is parametric, $r = r(j)$ is determined by \Eq{eq:j}, and $j = a^{-1/2}J/\hat{J}$, as defined above. Then, \Eq{eq:ndr0} can be rewritten as
\begin{gather}\label{eq:drpw}
\omega^2 = \omega_p^2\,\frac{2}{a}\int^\infty_0 g(j) F(J)\,dJ.
\end{gather}
Here $F(J)$ is the action distribution, and $g(j) \equiv [\pd_a\mc{H}(J, a)]/(mu^2)$, \ie $g(j(r)) = 2r - 1 - j(r)/j'(r)$,
\begin{gather}\label{eq:gasym}
g(j) = \left\{ 
\begin{array}{ll}
\ds -1 + j + \ldots, & \quad j \ll 1,\\[3pt]
\ds \frac{1}{2j^2} + \frac{5}{16j^6}+ \ldots, & \quad j \gg 1.
\end{array} 
\right.
\end{gather}
[Notice that $g(j)$ is continuous at the separatrix, with $g(j_*) = 1$, yet with a discontinuous infinite derivative; \Fig{fig:fig}(b).] In particular, when plasma is cold and $a \to 0$, then all particles are passing and $J = |p|/k$,~so
\begin{gather}\label{eq:vlin}
v^{\pm} = u \pm kJ/m,
\end{gather}
where the sign index denotes $\text{sgn}\,(v-u)$. Yet $v^+$ are not present then, and $v^{-} \ll u$, in which case \Eq{eq:vlin} gives $J \approx \hat{J}$ [in other words, one may assume $F(J) \approx \delta(J - \hat{J})$]. Since $g(j \gg 1) \approx 1/(2j^2)$, one thereby obtains $g = a/2$, meaning that \Eq{eq:drpw} predicts $\omega^2 = \omega_p^2$, as expected. 

Equation \eq{eq:drpw} describes \textit{all} kinetic corrections (to the extent that the monochromatic-wave approximation applies), and it readily shows how particles with given $j$ affect the wave frequency. In particular, it shows that deeply trapped particles ($j \lesssim 0.96$) reduce $\omega^2$, for the corresponding $g$ is negative; yet those near the separatrix and untrapped ones have positive $g$ and thus increase $\omega^2$ (cf. \Ref{ref:ikezi78}). Below, we explicitly calculate $\omega^2$ for a number of representative cases, by formally considering various asymptotic expansions of the integral in \Eq{eq:drpw}. 

\msection{Smooth distribution $F(J)$} First, let us assume that the distribution function $F(J)$ remains finite at small $J$ or, at least, diverges less rapidly than $J^{-1}$. Then, one can take the integral in \Eq{eq:drpw} by parts and obtain
\begin{gather}\label{eq:drpw2}
1 - \frac{2\omega^2_p}{a\omega^2} \int^\infty_0 \Psi(J, a) F'(J)\,dJ = 0,
\end{gather}
where we introduced $\Psi \equiv - \int^J_0 g(j)\,dJ = \hat{J}a^{1/2}\psi(j)$ and $\psi(j) \equiv - \int^j_0 g(\tilde{j})\,d\tilde{j}$ [\Fig{fig:fig}(c)], so
\begin{gather}\label{eq:psiasym}
\Psi (J, a) = \left\{ 
\begin{array}{ll}
\ds J - \frac{J^2}{2\hat{J}a^{1/2}} + \ldots, & \quad J \ll \hat{J}a^{1/2},\\[10pt]
\ds \frac{a \hat{J}^2}{2J} + \frac{a^3 \hat{J}^6}{16 J^5}+\ldots, & \quad J \gg \hat{J}a^{1/2}.
\end{array} 
\right.
\end{gather}

At $a \ll 1$, $\Psi$ changes rapidly with $J$ compared to $F(J)$, if the distribution is smooth, \eg thermal. Then, without using the explicit form of $\Psi (J, a)$ but rather drawing on the leading terms in \Eq{eq:psiasym}, one can put \Eq{eq:drpw2} in the following asymptotic form:
\begin{gather}\label{eq:ndr}
\epsilon(\omega, k)  + \frac{\omega_p^2}{2k^2}\,C_1 \ln a + \frac{\omega\omega_p^2}{k^3}\,\varkappa C_2a^{1/2} = 0.
\end{gather}
Here we introduced
\begin{gather}\label{eq:epsilonF}
\epsilon = 1 -\frac{m^2\omega_p^2}{k^6} \int^{\infty}_0 \left[F'(J) - F'(0)\,q\!\left(\frac{J}{\hat{J}}\right)\right]\,\frac{dJ}{J},
\end{gather}
$q(z) = 1 - 2z\psi(z)$ [\Fig{fig:fig}(d)], $C_1 = (m/k)^2F'(0)$, $C_2 = (m/k)^3F''(0)$, and $\varkappa = \int^\infty_0 q(z)\,dz  \approx 0.544$. [Notice that, albeit determined by essentially nonlinear dynamics in the narrow vicinity of the resonance, $q(z)$ nevertheless can affect the integrand on the thermal scale.] In particular, when $F'(0) = 0$, the nonlinear part of \Eq{eq:ndr} is small, yielding that the nonlinear frequency shift $\Delta \omega$ is also small; hence, 
\begin{gather}\label{eq:dwlin}
\Delta \omega = - \left(\frac{\pd \epsilon}{\pd \omega}\right)^{\!\!-1}\!\frac{\varkappa\omega_p^2}{k^2}\sqrt{\frac{e\varphi_0}{m}}\,C_2.
\end{gather}
Yet, at nonzero $F'(0)$, the nonlinear part of \Eq{eq:ndr} diverges logarithmically at small $a$; \ie wave interaction with resonant particles has a strong effect on $\omega$.

Equations \eq{eq:ndr}-\eq{eq:dwlin} generalize the existing NDR for eigenwaves in plasmas with smooth distributions \cite{ref:dewar72b, refm:sqrt, ref:lindberg07}, namely, as follows. First of all, notice that $\epsilon$, serving as a \textit{generalized linear dielectric function} here, is a functional of the action distribution. Unlike the commonly used distribution of ``unperturbed'' velocities $f_0(v)$, which depends on how the wave was excited \cite{ref:dewar72b}, $F(J)$ is defined unambiguously; thus, the above equations hold for any excitation scenario [while finding $F(J)$ itself is kept as a separate problem]. Second, even if put in terms of $f_0(v)$, \Eqs{eq:ndr}-\eq{eq:dwlin} cover a wider class of particle distributions. The latter is seen as follows.

For example, consider a wave developed slowly from zero amplitude, so each $J$ is conserved, even through trapping and untrapping \cite{ref:best68, refm:separatrix, ref:lindberg07}. Then $F(J) = F_0(J)$, index 0 henceforth denoting the initial state. Yet, since there was no wave in that state, \Eq{eq:vlin} applies, so each $\ell$th derivative of $F_0(J)$ reads as $F^{(\ell)}_0(J) = (k/m)^\ell [f_0^{(\ell)}(v^{+}) + (-1)^\ell f_0^{(\ell)}(v^{-})]$. Let us use bars to denote limits $f_0^{(\ell)}(v \to u \pm)$, so that $\bar{f}_0^{(\ell)}(v)$ is defined as a piecewise-constant function equal to the left and right limits for $v < u$ and $v > u$ correspondingly. Then, $C_1 = \bar{f}'_0(u+) - \bar{f}'_0(u-)$, $C_2 = \bar{f}''_0(u+) + \bar{f}''_0(u-)$, and
\begin{gather}\label{eq:eplin}
\epsilon = 1 -\frac{\omega_p^2}{k^2} \int^{\infty}_{-\infty} \,\frac{f_0'(v) - q\bar{f}'_0(v)}{v-u}\,dv,
\end{gather}
where $q \equiv q(|v/u-1|)$. (Remarkably, contributing to $\epsilon$ are both passing and trapped particles.)

First, compare \Eq{eq:eplin} with the usual $\epsilon_L = 1 -(\omega_p^2/k^2)\, \msf{P}\int^{\infty}_{-\infty} (v-u)^{-1}f_0'(v)\, dv$, $\msf{P}$ denoting the principal value \cite{book:stix}. For smooth $f_0(v)$, our $\epsilon$ can be put in the same form as $\epsilon_L$, because $\msf{P}\int^{\infty}_{-\infty} (v-u)^{-1}q\bar{f}'_0(v)\, dv = 0$. However, \Eq{eq:eplin} is valid also when $f'_0(v)$ is discontinuous across the resonance, a case in which $\epsilon_L$ is undefined. This is because the integrand in \Eq{eq:eplin} is finite (piecewise-continuous), so the integral converges absolutely rather than existing only as a principal value (like $\epsilon_L$ does). Second, for smooth $f_0(v)$, when $C_1 = 0$ and $C_2 = 2f''_0(u)$, \Eq{eq:dwlin} for $\Delta \omega$ matches that in \Ref{ref:dewar72b}, including the coefficient. Yet, unlike the existing theory, our \Eqs{eq:ndr}-\eq{eq:eplin} apply just as well for arbitrary $C_1$ and $C_2$, in which case $f_0(v)$ may not be smooth while $F(J)$~is. 

\msection{Beam nonlinearities} Suppose now that, in addition to a smooth distribution $\mc{F}(J)$, near the resonance there is a phase-space clump or a hole, further termed uniformly as a beam with ${F_b(J) \gtrless 0}$ and some average spatial density ${n_b \gtrless 0}$; namely, $F(J) = \mc{F}(J) + F_b(J)$. For example, take $F_b(J) = (n_b/n_0)\delta(J)$, where $n_0$ is the bulk density that enters here due to normalization. Since $g(0) = - 1$, \Eq{eq:drpw} yields then, with $\omega_b^2 = 4\pi n_be^2/m$:
\begin{gather}\label{eq:beamdisp}
\omega^2 = \omega_L^2 - 2\omega_b^2/a,
\end{gather}
[here nonlinearities due to $\mc{F}$ are neglected, and $\omega^2 = \omega_L^2(\omega, k)$ corresponds to the linear equation], or, more specifically, $\epsilon(\omega, k) + 2\omega_b^2/(a\omega^2) = 0$. These equations agree with the known NDR for modes with deeply trapped particles \cite{ref:krasovsky94, refm:deep} and leads to $\Delta \omega = \mc{O}(a^{-1})$, such that $\Delta \omega < 0$ for a clump and $\Delta \omega > 0$ for a hole. 

Now let $n_b$ itself depend on the wave amplitude. For example, a Van Kampen mode would have $2\omega_b^2/(a\omega_L^2) \equiv \eta$ of order one \cite{foot:eta}; in this case, by adjusting $a$, any $\omega$ can be produced for a given $k$, in agreement with the linear theory \cite{ref:vankampen55, book:stix}. Also, consider the case when $F_b$ is constant across the trapping width: $F_b(J) = F_b\Theta(J) \Theta(J_* - J)$, with $\Theta$ being the Heaviside step function. Then $n_b$ is proportional to the separatrix action $J_*$ [\ie $n_b = \mc{O}(a^{1/2})$], and one gets
\begin{gather}\label{eq:aux2}
\omega^2 = \omega_L^2 - [8/(3\pi)]a^{-1/2}\omega_p^2\hat{J}F_b,
\end{gather}
since $\psi(j_*) = 4/(3\pi)$. Equation \eq{eq:aux2} also matches the result found previously, \eg in \Ref{ref:krasovsky07}. 

Finally, consider dissipation-driven effects in collisional plasmas. Since $\omega$ changes rapidly with small $a$ in the presence of a phase-space clump or a hole, slow decay of $a$ will cause frequency downshifting or upshifting, correspondingly. Yet, since the power index $\sigma$ in the scaling $\Delta\omega \propto a^{-\sigma}$ depends on how localized $F_b(J)$ is, another effect is anticipated, namely, as follows. Notice that a friction drag (say, proportional to the particle velocity) can cause condensation of the trapped distribution near the bottom of the wave potential trough \cite{my:mefffric}. Hence, peaking of $F(J)$ can occur, and $\sigma$ can increase gradually up to unity. This represents a frequency sweeping mechanism additional to those considered in \Refs{refm:breizman}.

\msection{Conclusions} In summary, we show here that knowing the appropriate single-particle OC Hamiltonian is sufficient to derive the fully nonlinear dispersion of a stationary wave in collisionless plasma without solving Vlasov or Maxwell's equations. We illustrate how our theory reduces to results previously known from separate contexts, recovering them within a single NDR. In particular, for longitudinal electron oscillations in nonmagnetized plasma, various types of kinetic nonlinearities are derived, including a new logarithmic nonlinearity, simply by substituting appropriate distributions $F(J)$ into \Eq{eq:drpw}. Also, the linear dielectric function is generalized, and the friction drag on trapped particles is predicted to affect the wave frequency sweeping in collisional plasmas.

The work was supported through the NNSA SSAA Program through DOE Research Grant No. DE274-FG52-08NA28553.


\begin{thebibliography}{10}
\bibitem{foot:disp}
The inverse argument, that $\mc{H}$ is derivable from NDR, was discussed
  recently in \Refs{my:kchi, my:nlinphi}. Also see references therein
  pertaining to the OC approach in general.
\bibitem{foot:hl}
$L_i$ that enters \Eq{eq:ham} is determined by the wave spatial profile, which
  itself must be found. For anharmonic oscillations, $L_i$ and $\mc{H}_i$ can
  be constructed iteratively, along the lines of existing iterative approaches
  to NDR \cite{ref:rose01, ref:lindberg07}. Yet, in practice, the
  spatial profile is most often assumed to be prescribed (\eg sinusoidal). Within
  this commonly accepted approach, our method becomes particularly useful,
  because then $L_i$ is known and thus $\mc{H}_i$ is constructed deductively
  and nonperturbatively in~$a$.
\bibitem{foot:eta}
Strictly speaking, though, for \Eq{eq:beamdisp} to hold, required is the
  sinusoidal-wave approximation and thus $\eta \ll 1$ \cite{ref:krasovsky94}.
\end{thebibliography}
\end{document}